\documentclass[3p,times,twocolumn,number]{elsarticle}

\makeatletter
\let\c@author\relax
\makeatother

\usepackage[numbers]{natbib}
\usepackage{blindtext}
\usepackage{paralist}
\usepackage{amssymb}
\usepackage{color}
\usepackage{graphicx}
\usepackage[acronym]{glossaries}
\newacronym{api}{API}{Application Programming Interface}
\newacronym{apt}{APT}{Advanced Packaging Tool}
\newacronym{bom}{BOM}{Bill of Materials}
\newacronym{cgi}{CGI}{Common Gateway Interface}
\newacronym{cip}{CIP}{Cybersecurity Policy for Critical Infrastructure Protection}
\newacronym{cisa}{CISA}{Cybersecurity and Infrastructure Security Agency}
\newacronym{cla}{CLA}{Contributor License Agreement}
\newacronym{cna}{CNA}{CVE Numbering Authority}
\newacronym{cra}{CRA}{Cyber Resilience Act}
\newacronym{cve}{CVE}{Common Vulnerabilities and Exposures}
\newacronym{cvs}{CVS}{Concurrent Versions System}
\newacronym{dhcp}{DHCP}{Dynamic Host Configuration Protocol}
\newacronym{dns}{DNS}{Domain Name System}
\newacronym{dos}{DOS}{Denial of Service}
\newacronym{ecn}{ECN}{Explicit Congestion Notification}
\newacronym{fda}{FDA}{Food and Drug Administration}
\newacronym{fdc}{FD\&C Act}{Federal Food Drug and Cosmetic Act}
\newacronym{foss}{FLOSS}{Free/Libre and Open Source Software}
\newacronym{http}{HTTP}{Hypertext Transfer Protocol}
\newacronym{ietf}{IETF}{Internet Engineering Task Force}
\newacronym{iqr}{IQR}{Inner Quartile Range}
\newacronym{jndi}{JNDI}{Java Naming and Directory Interface}
\newacronym{json}{JSON}{JavaScript Object Notation}
\newacronym{ldap}{LDAP}{Lightweight Directory Access Protocol}
\newacronym{lma}{LMA}{Level of Maintenance Activity}
\newacronym{loc}{LoC}{lines of code}
\newacronym{nist}{NIST}{National Institute of Standards and Technology}
\newacronym{npo}{NPO}{Nonprofit organization}
\newacronym{nvd}{NVD}{National Vulnerability Database}
\newacronym{ossf}{OSSF}{Open Source Security Foundation}
\newacronym{owasp}{OWASP}{Open Web Application Security Project}
\newacronym{p2p}{P2P}{Peer-to-Peer}
\newacronym{pypi}{PyPI}{Python Package Index}
\newacronym{rce}{RCE}{Remote Code Execution}
\newacronym{sbom}{SBOM}{Software Bill of Materials}
\newacronym{spdx}{SPDX}{System Package Data Exchange}
\newacronym{tcp}{TCP}{Transmission Control Protocol}
\newacronym{tls}{TLS}{Transport Layer Security}
\newacronym{udp}{UDP}{User Datagram Protocol}
\newacronym{url}{URL}{Uniform Resource Locator}
\newacronym{vcs}{VCS}{Version Control System}
\newacronym{osv}{OSV}{Open Source Vulnerabilities Database}

\usepackage[hidelinks]{hyperref}
\usepackage{array}
\usepackage{booktabs}
\usepackage{cleveref}
\usepackage{pifont}
\usepackage{arydshln}
\usepackage{orcidlink}
\usepackage{fancyvrb}
\usepackage{caption}
\usepackage{subcaption}
\usepackage{xurl}

\usepackage[english]{babel}
\addto\extrasenglish{%

}

\newcommand{\RQ}[0]{“How can problematic parts in the \gls{foss} ecosystem be identified?”}
\newcommand{\RQQ}[0]{“What is the current state of the \gls{foss} ecosystem?”}

\journal{Future Generation Computer Systems}

\begin{document}
\begin{frontmatter}

	\title{Tracking Down Software Cluster Bombs: A Current State Analysis of the Free/Libre and Open Source Software (FLOSS) Ecosystem}

	\author[1,2]{\orcidlink{0000-0002-2288-9010} Stefan Tatschner\corref{cor1}}
	\author[1,3,4]{\orcidlink{0000-0002-1094-4828} Michael P. Heinl}
	\author[2]{\orcidlink{0009-0008-0767-8208} Nicole Pappler}
	\author[1]{\orcidlink{0009-0001-7615-7579} Tobias Specht}
	\author[5]{\orcidlink{0000-0002-1658-1140} Sven Plaga}
	\author[2]{\orcidlink{0000-0002-3375-8200} Thomas Newe}

	\cortext[cor1]{Corresponding author}

	%
	%

	%
	%
	%

	\affiliation[1]{organization={Fraunhofer AISEC},
		city={Garching bei München},
		state={Bavaria},
		country={Germany}}
	\affiliation[2]{organization={University of Limerick},
		city={Limerick},
		addressline={V94 T9PX},
		country={Ireland}}
	\affiliation[3]{organization={Technical University of Munich},
		city={Garching bei München},
		state={Bavaria},
		country={Germany}}
	\affiliation[4]{organization={Munich University of Applied Sciences HM},
		city={Munich},
		state={Bavaria},
		country={Germany}}
	\affiliation[5]{organization={Center for Intelligence and Security Studies (CISS)},
		city={Neubiberg},
		state={Bavaria},
		country={Germany}}

	\begin{abstract}

		Throughout computer history, it has been repeatedly demonstrated that critical software vulnerabilities can significantly affect the components involved.
		In the \gls{foss} ecosystem, most software is distributed through package repositories.
		Nowadays, monitoring critical dependencies in a software system is essential for maintaining robust security practices.
		This is particularly important due to new legal requirements, such as the European Cyber Resilience Act, which necessitate that software projects maintain a transparent track record with \gls{sbom} and ensure a good overall state.
		This study provides a summary of the current state of available \gls{foss} package repositories and addresses the challenge of identifying problematic areas within a software ecosystem.
		These areas are analyzed in detail, quantifying the current state of the \gls{foss} ecosystem.
		The results indicate that while there are well-maintained projects within the \gls{foss} ecosystem, there are also high-impact projects that are susceptible to supply chain attacks.
		This study proposes a method for analyzing the current state and identifies missing elements, such as interfaces, for future research.
	\end{abstract}


	\begin{highlights}
		\item Methodology for finding problematic parts in a software ecosystem
		\item Conducted an analysis of a Linux distribution's current state in terms of software dependencies
		\item Analysis of the practicability of new legal requirements
	\end{highlights}

	\begin{keyword}
		Dependency Graph, Vulnerability Databases, Centrality, Software Supply Chain Defects
	\end{keyword}

\end{frontmatter}

\section{Introduction}
\subsection{Motivation}


Throughout computer history, there are numerous instances where fundamental components of the software ecosystem have been affected by critical vulnerabilities, such as \gls{rce} or Information Disclosure.
It is inherent that insecure shared components cause repercussions for every software component that relies on the vulnerable component.
The notorious Heartbleed bug (CVE-2014-0160) was a software vulnerability that impacted a vast number of servers.
This was because the affected component, OpenSSL, underpins most encrypted internet traffic.
The Heartbleed vulnerability is still remembered to this day.


Many software components, such as email, web, or database servers, require an implementation of \gls{tls}~\cite{rfc8446}.
Today, OpenSSL provides a maintained and current implementation of \gls{tls}, allowing applications to rely on OpenSSL rather than re-implementing the protocol.
As a result, a critical vulnerability in OpenSSL is likely to affect the entire ecosystem.
For example, at the time of Heartbleed's disclosure, approximately 300,000 vulnerable servers were online.
Six years later, about 200,000 vulnerable servers remained online \cite{Morris2024}.
Other vulnerabilities of shared basic components that also drew significant media attention include Ghost (CVE-2015-0235), Log4Shell (CVE-2021-33228), and the attempt to insert a backdoor into the xz compression library (CVE-2024-3094).


Recently, an \gls{rce} vulnerability (CVE-2023-4863) was discovered in the widely used image decoding library \texttt{libwebp}.
This library is employed for decoding the WebP image format in Google Chrome and Firefox.
According to an analysis by Google, every user of \texttt{libwebp} was affected by this vulnerability.
This raises the question: “Which software components or products are affected?“
In other words, which software components depend on the vulnerable module, and which could have repercussions on a significant portion of the software ecosystem?


From a technical standpoint, sharing code in the form of libraries is highly beneficial, particularly when these libraries implement security-related features \cite{PLAGA2019596}.
This approach allows for the combination of efforts and helps avoid recurring problems or anti-patterns~\cite{heinl2020} by centralizing relevant code paths.
In practice, there exists a diverse collection of libraries with different approaches.
For example, well-known \gls{p2p} applications share and expose their underlying networking techniques to be reused by other applications~\cite{s21154969}.
In contrast, new technologies such as the QUIC protocol often undergo multiple implementations until one proves itself effective in practice~\cite{10.1145/3600160.3605164}.

\subsection{Problem Statement}\label{sec:problemStatement}


These various approaches result in a situation where a few key components, which perform basic yet important tasks, become essential for many applications, as there are often no alternatives available.
Often, these key components are maintained by a few or even a single developer, who may work on them during their free time rather than as part of their employment.
These components can significantly impact the overall ecosystem and should be maintained with particular care.

\subsection{Paradigmatic Vulnerabilities}\label{sec:paradigmaticVulns}

The described situation has repeatedly led to severe vulnerabilities in the past.
The following examples demonstrate that a critical vulnerability in a reused library can have repercussions on many other components:

\begin{itemize}
	\item \textbf{Heartbleed (CVE-2014-0160)}: This vulnerability in OpenSSL led to information leaks of sensitive data.
	      Before 2014, OpenSSL suffered from poor code quality, presumably due to internal project issues such as the absence of testing or code reviews, resulting from insufficient funding.
	      Major software projects, such as nginx, postfix, and CPython, depend on OpenSSL.
	\item \textbf{Shellshock (CVE-2014-6271)}: This vulnerability (family) in Bash caused privilege escalation and \gls{rce}.
	      According to its Git repository, Bash appears to be primarily maintained by a single person; its Git repository contains only \emph{two} different authors.
	      Because Bash is a central component of most Linux systems, multiple services were affected by this vulnerability, including web servers based on \gls{cgi}, \gls{dhcp} clients, and OpenSSH.
	\item \textbf{Ghost (CVE-2015-0235)}: A buffer overflow bug affecting the \texttt{gethostbyname()} and \texttt{gethostbyname2()} function calls in the \texttt{glibc} library.
	      This vulnerability allows a remote attacker to execute arbitrary code with the permissions of the user running the application.
	      Since \texttt{glibc} is a very basic library that provides programming interfaces to communicate with the underlying operating system, it is used by almost all software modules.
	      Consequently, a large number of programs were affected by Ghost.
	\item \textbf{Log4Shell (CVE-2021-33228)}: This vulnerability had existed unnoticed in the Log4j logging framework since 2013.
	      The vulnerability exploits Log4j's ability to allow requests to arbitrary \gls{ldap} and \gls{jndi} servers, enabling attackers to execute arbitrary Java code.
	      The exploit is estimated to have had the potential to affect hundreds of millions of devices~\cite{log4j}.
	\item \textbf{The WebP 0day (CVE-2023-4863)}: With a specially crafted WebP lossless file, \texttt{libwebp} may write data out of bounds to the heap.
	      Attacks against this vulnerability can range from \gls{dos} to possible \gls{rce}.
	      WebP is widely used in web applications, thus primarily affecting browsers or email clients such as Google Chrome, Firefox, and Thunderbird.
	\item \textbf{The xz Backdoor (CVE-2024-3094)}: The widely used xz compression library suffered from a supply chain attack attempting to insert a master key for large-scale root access.
	      A malicious actor gained trust and maintainer access to the project's source code over time.
	      Eventually, highly sophisticated backdoor code was added to the repository, disguised as test files.
	      Through its legacy autotools-based build system, the malicious code was incorporated into the resulting shared library\footnote{\url{https://research.swtch.com/xz-script}}.
	      The added code was used to overwrite the authentication routines of the SSH service at runtime by exploiting a GCC feature related to dynamic linking.
\end{itemize}

\subsection{Legal Regulations}


When software vulnerabilities in third-party libraries are discovered, vendors using these libraries in their products are pressured to fix them as quickly as possible, since they may be disclosed and exploited by attackers.
Identifying and capturing detailed software composition information, including transitive dependencies, is therefore an essential tool for monitoring risks in the software supply chain.


ISO/IEC 5230/2020 \cite{ISO5230}, in combination with ISO/IEC 18974:2023 \cite{ISO18974}, known as OpenChain and the OpenChain Security Assurance Specification, define structures and principles to maintain control over inbound and outbound software.
Two primary standards for creating \glspl{sbom} have been established so far:

\begin{itemize}

	\item \textbf{\gls{spdx}}\footnote{\url{https://spdx.dev/}}, a project initiated by the Linux Foundation, which also became ISO/IEC 5962:2021 \cite{ISO5962}, and
	\item \textbf{CycloneDX}\footnote{\url{https://cyclonedx.org/}}, a format developed by the \gls{owasp} community.
\end{itemize}
%

\glspl{sbom} created according to \gls{spdx} and CycloneDX both provide machine-readable formats, allowing for the analysis of whether a particular software application is affected by a newly known vulnerability.


In order to track security vulnerabilities, the \gls{cve} standard was published by the MITRE Corporation\footnote{\url{https://cve.mitre.org/}}, in collaboration with US government agencies.
Although not directly required by any current regulation, the \gls{cve} system has been widely adopted by the cybersecurity community.
It is widely used by tools and software distributors, becoming the de facto method for referring to software vulnerabilities.
As security incidents become more frequent and sophisticated, governments are starting to introduce regulatory requirements addressing supply chain issues.


In the US, the government explicitly promotes \glspl{sbom} through an executive order\footnote{\url{https://www.whitehouse.gov/briefing-room/presidential-actions/2021/05/12/executive-order-on-improving-the-nations-cybersecurity}}.
The \gls{fda} requires the provision of \glspl{sbom} with all medical devices in its Medical Device Cybersecurity amendment to the \gls{fdc}\footnote{\url{https://www.fda.gov/media/119933/download}}, and the most recent Cybersecurity Framework\footnote{\url{https://www.nist.gov/cyberframework}} published by the \gls{nist} requires a Supply Chain Risk Management program including inventories of hardware and software.
Japan recently published the latest version of its \gls{cip}\footnote{\url{https://www.nisc.go.jp/eng/pdf/cip_policy_2024_eng.pdf}}, calling for risk management to address attacks originating from the supply chain.
The European \gls{cra}\footnote{\url{https://www.europarl.europa.eu/doceo/document/TA-9-2024-0130_EN.html}}, adopted on March 12, 2024 by the European Parliament, is the latest addition to these national and international regulations, explicitly requesting \glspl{sbom}.


As the number of dependencies listed in \gls{sbom} documents can be very high, managing these dependencies will become a crucial part of establishing effective measures in a cybersecurity governance process.

\subsection{Research Questions}


Based on the increasing legal requirements, this study aims to provide a high-level view of the \gls{foss} ecosystem.
The basic concept of \glspl{sbom} and the legal demand for tracking software dependencies might generally be a good idea.
However, \emph{identifying} issues in the software supply chain via \glspl{sbom} does not address the root cause of the inherent problem.
To \emph{avoid} such issues, the first step is to understand the current state of the \gls{foss} ecosystem.
Such an overview will help identify parts of the supply chain where, for example, financial support or human resources are needed.
For instance, the recent xz Backdoor could potentially have been avoided with better funding for the maintainer from the outset\footnote{\url{https://research.swtch.com/xz-timeline}}.
These considerations lead to research question \textbf{RQ1} addressed by this paper: \RQ.
Extending this idea further, a more generalized research question \textbf{RQ2} can be formulated: \RQQ.


Previous research has shown that software ecosystems are structured similarly to social networks \cite{9631870,7490780}, thus relevant methods and potential data sources from this area are examined.
The NixOS \cite{10.1145/1411203.1411255} ecosystem is of interest because it uses the Nix programming language to describe software packages and their relationships.
NixOS presents several compelling advantages that render it particularly suitable for scientific research.
Its extensive package repository surpasses that of Debian with over 80,000 packages.
Furthermore, NixOS's dependency management system, characterized by its functional package management, ensures precise and reproducible handling, and definition of dependencies.
Additionally, the monorepo structure of NixOS consolidates all package definitions into a single repository, thereby eliminating the need to crawl and parse multiple sources, which significantly streamlines the process of package analysis and data extraction.
These attributes collectively establish NixOS as an ideal platform for researchers seeking a reliable, comprehensive, and efficient system for conducting software-based scientific investigations.


The presented paper proposes a method for identifying problematic parts of the \gls{foss} ecosystem, as described in \autoref{sec:problemStatement}.
This method draws on established techniques from graph theory and related areas of study.
The developed method is applied to a real-world software package repository where metrics known from the literature, such as project age or the number of lines of code, are collected and evaluated.
This paper provides insights and related methods to identify critical projects.
Based on the results of this paper and new legal requirements for dependency tracking, there is potential for future research in this field.

\section{Related Work}


The identified related work can be organized into three distinct categories.
For each category, a literature review was conducted to represent the theoretical background of the analysis presented in this paper.


\begin{itemize}
	\item \textbf{Centrality in Graphs}: This section covers the theoretical background of the selected graph-based methods. The research focuses on the applicability of established evaluation methods known from social networks.
	\item \textbf{Package Dependency Analysis}: This section examines previous studies related to the structure of software ecosystems like the npm Registry, RubyGems.org, or the Debian project.
	\item \textbf{Project Status Analysis}: Research is conducted on techniques for quantifying the maintenance status of software projects.
\end{itemize}

\subsection{Centrality in Graphs}


In the context of network analysis, centrality measures offer several advantages over naive methods, such as merely counting incoming edges.
Unlike these naive approaches, centrality provides a nuanced understanding of a node's importance based on its position within the network.
By employing centrality, one can gain insights into the overall network structure, including identifying key influencers, potential bottlenecks, or vulnerable nodes.
Overall, centrality offers a more comprehensive view of node importance within a network compared to simply counting incoming edges.


Centrality algorithms assign numbers or rankings to nodes within a graph based on their network position.
Applications include identifying the most influential people in a social network, key infrastructure nodes on the internet, or analyzing urban networks.
In general, centrality algorithms answer the question, “What characterizes an important node?”
The word “importance” can have many meanings, making the available definitions of centrality versatile \cite{Freeman1978}.


Among the various centrality algorithms, eigenvector centrality emerged as the most promising algorithm for this study, as demonstrated in 2019 by Gómez \cite{Gomez2019}.
While eigenvector centrality focuses on the importance of connections to influential nodes, betweenness centrality highlights nodes that facilitate communication between others, and closeness centrality emphasizes nodes with efficient access to the entire network.
Other centrality algorithms were deemed unsuitable for this study.


Historically, eigenvector centrality was introduced by Landau \cite{landau1895relativen} for chess tournaments.
Half a century later, it was rediscovered by Wei \cite{wei1952algebraic} and popularized by Kendall \cite{760e07d1-fd0d-3ce0-afae-f7ab9cd57766} in the context of sports ranking.
Claude introduced a general definition for graphs based on social connections \cite{claude1966theorie}.
Eventually, Bonacich \cite{35397813-90c1-3806-8d5d-a07b3340ac3d} reintroduced eigenvector centrality and made it popular in link analysis.


Eigenvector centrality measures the influence of a node based on the connections of the nodes to which it is connected.
Similar to degree centrality, eigenvector centrality favors nodes with a high number of links.
Unlike degree centrality, eigenvector centrality also considers the centrality of the adjacent nodes.


Due to its mathematical foundation, eigenvector centrality requires strongly connected graphs\footnote{\url{https://ocw.mit.edu/courses/14-15-networks-spring-2022/mit14_15s22_lec3.pdf}}.
In the context of a software repository's dependency graph for a Linux distribution, strong connectivity cannot always be assumed.
However, a more general variant exists: the Katz centrality algorithm, introduced by Leo Katz in 1953 \cite{Katz1953}.
Unlike eigenvector centrality, Katz centrality also applies to graphs that are not strongly connected.


Katz centrality is capable of assigning scores to nodes outside the largest connected component.
It incorporates an attenuation factor to account for paths of varying lengths, ensuring non-zero scores for a broader range of nodes.
By emphasizing immediate neighbors through a constant additive term, Katz centrality considers both direct and indirect connections.
This robustness benefits nodes with fewer connections that remain influential due to their network positions.
The attenuation factor provides flexibility, allowing for adjustments based on the network's unique characteristics.
Overall, Katz centrality offers a versatile approach for evaluating disconnected networks, delivering meaningful scores across the graph.

\subsection{Package Dependency Analysis}


In 2015, Wang et al. published a study using a graph-based method to create a distribution-wide dependency analysis for Ubuntu~14.04 \cite{7490780}.
The authors present the challenges of creating a dependency graph by parsing package metadata from package managers such as Debian's \gls{apt}.
This work illustrates that a graph-based approach is efficient for understanding the software structure of an entire distribution and can assist in further, more detailed investigations.


In 2017, Decan, Mens, and Claes published a comparison of dependency issues in \gls{foss} packaging ecosystems \cite{7884604}.
The authors presented an empirical analysis of the evolution of dependency graphs of three large package ecosystems.
The paper highlights solutions each package ecosystem has implemented for dependency update issues, such as dependency constraints.
The authors conclude that package dependency updates entail a non-negligible maintenance cost and that better packaging and dependency analysis tools are needed.


In 2018, Decan, Mens, and Grosjoen published a more detailed study on the evolution of software packaging ecosystems \cite{Decan2019}.
The authors state that most packages depend on other packages, but only a small proportion of packages account for most reverse dependencies.
According to the study, there is a high proportion of so-called “fragile” packages due to a high and increasing number of transitive dependencies over time.
The study concludes that these findings are instrumental for assessing the quality of package dependency networks and that improvements can be made through more comprehensive dependency management tools and imposed policies.


In 2021, Suhaib Mujahid et al. published a study evaluating a centrality-based approach that could detect packages in the npm Registry that are in decline~\cite{9631870}.
The authors conclude that it is possible to predict when packages in the npm ecosystem will soon become deprecated.
The article's key point is an analysis of the popular npm package Moment.js. The authors published a chart showing a declining centrality value since September 2018.
Two years later, the package was considered deprecated by its developers.
It was not until that point in time that the number of packages depending on Moment.js began to drop.


In 2025, Alhamdan and Staicu published a study analyzing the Deno ecosystem \cite{Alhamdan:Staicu:2025}.
The authors state that although Deno has a smaller attack surface than Node.js, several attacks are not addressed or only partially addressed.
The paper also highlights that classical URL-related issues, such as expired domains or reliance on insecure transport protocols, remain relevant.
The authors propose the following improvements to the security model of the Deno ecosystem: add import permissions, additional access control at the file system level, support for compartmentalization, and a manifest file that persists fine-grained permissions.

\subsection{Project Status Analysis}


Unmaintained projects continue to pose a serious problem, as also shown by \cite{236368}.
In 2020, Jailton Coelho et al. published a study \cite{COELHO2020106274} proposing a method to identify GitHub projects that are not actively maintained.
The authors introduced the so-called \gls{lma} value as a metric for describing the maintenance status of GitHub projects.
They trained a machine learning model to identify unmaintained or sparsely maintained projects based on a set of features, such as the number of commits, forks, or issues.
The approach was released as an extension for Google Chrome.
Ironically, this extension is no longer actively maintained.
Furthermore, the published extension does not contain the trained model but instead is intended to communicate with a server provided by the authors.
This server is currently offline, and therefore the developed approach cannot be used for this paper.
However, the authors conducted several correlation analyses.
They state that factors such as the number of contributors or lines of code can be used to assess the maintenance status of a software project, as these are correlated with their \gls{lma} value.


In 2020, Rob Pike published a technical article \cite{pike2020} describing the criticality score, a technique for quantifying criticality.
In 2021, this technique was further examined by Pfeiffer \cite{pfeiffer}.
The goal of this score is to find a single value that meaningfully represents all signals of criticality for a package.
The \gls{ossf} group, operated by the Linux Foundation, maintains multiple software projects on GitHub\footnote{\url{https://github.com/ossf}} that can be used to calculate the criticality score of GitHub projects or conduct surveys about the state of software ecosystems.
However, their tools are only applicable within a limited scope.
For instance, the official criticality score tool\footnote{\url{https://github.com/ossf/criticality_score}} is only applicable to GitHub projects, and a Google Cloud account is required to use the tool.

\section{Methodology}
\label{sec:methodology}


This research consists of multiple, consecutive working steps.
The results were collected in separate tables, including additional metadata such as the location of Git repositories or further references.
As these tables form the database for the evaluation, they will be published as supplementary data to this paper.
From a conceptual point of view, the methodology is structured as follows:
\begin{enumerate}
	\item \textbf{Identification of relevant packages}: Relevant packages were identified using a graph-based approach. The nixpkgs\footnote{\url{https://github.com/NixOS/nixpkgs}} repository was used to create a dependency graph of all (at the time of writing) 82,011 software packages.
	      Unlike similar repositories, e.g., npm, PyPi, or crates.io, the nixpkgs repository also contains information about system dependencies.
	      Analyzing alternative software repositories would yield ecosystem-specific results and exclude system libraries such as libssl (provided by OpenSSL) or libcurl (provided by the curl project) from the analysis.
	      The authors considered the nixpkgs repository suitable for this study since no technical limitations exist for creating a full dependency graph including system dependencies.

	      The nixpkgs repository is built on the special nix language, which allows building a large graph data structure without needing to parse text-based metadata.
	      Subsequently, techniques developed in the field of social network analysis, also known as \emph{indicators of centrality}, were used to determine the 200 most important nodes in the dependency graph for subsequent manual analysis.
	      Some packages were not relevant for further examination, as they, e.g., contained only documentation.
	      After manual review, 35 packages were filtered out.
	      The graph library NetworkX \cite{SciPyProceedings_11} was used for graph processing.

	\item \textbf{Identification of relevant vulnerability databases}: The NixOS project does not maintain a vulnerability tracker where actual \glspl{cve} are mapped to real package names and their state is tracked. For this study, the Debian vulnerability database\footnote{\url{https://security-tracker.debian.org/tracker/}} was used as a data source for CVE information. Other large databases, such as OSV\footnote{\url{https://osv.dev/}} maintained by Google \cite{10.1145/3597503.3639582}, were considered unsuitable as there is no mapping of CVEs to actual software packages available.
	\item \textbf{Addition of missing data}: The identified packages were reviewed, and missing data, such as the location of the Git repository, category, license, or implementation language, were added manually to the table.
	\item \textbf{Collection of metrics}: The Git repository of every identified software package was cloned and examined. Multiple values were extracted, such as the number of contributors, age, or commit activity. These values were added to the table as well.
	\item \textbf{Collection of \glspl{cve}}: The package IDs from the nixpkgs repository were manually mapped to the corresponding package IDs in the Debian repository. The Debian security database was downloaded, and the \gls{cve} stats were mapped to the relevant nixpkgs packages.
	\item \textbf{Analysis of gathered data}: The data was evaluated and visualized using well-known methods from data science, such as scatter plots, bar graphs, or pie charts.
\end{enumerate}

\section{Data Gathering}

\subsection{Determination of Relevant Packages}


In graph theory and network analysis, indicators of centrality assign numbers or rankings to nodes within a graph based on their network position.
Applications include identifying the most influential nodes in social networks, computer networks, or urban networks.
Centrality is even applicable in identifying super-spreaders of diseases.
A high Katz centrality score indicates strong influence over other nodes in the network.
It is useful because it signifies not just direct influence, but also influence over nodes more than one hop away.


From the literature research, Katz centrality was found to be applicable to a dependency graph data structure describing software dependencies.
Practically, the centrality algorithm assigns each node a numerical score.
Finally, this score can be used to sort the nodes (i.e., the software packages) according to their importance in the software ecosystem.


\begin{figure}[htb]
	\centering
	\includegraphics[width=0.8\linewidth]{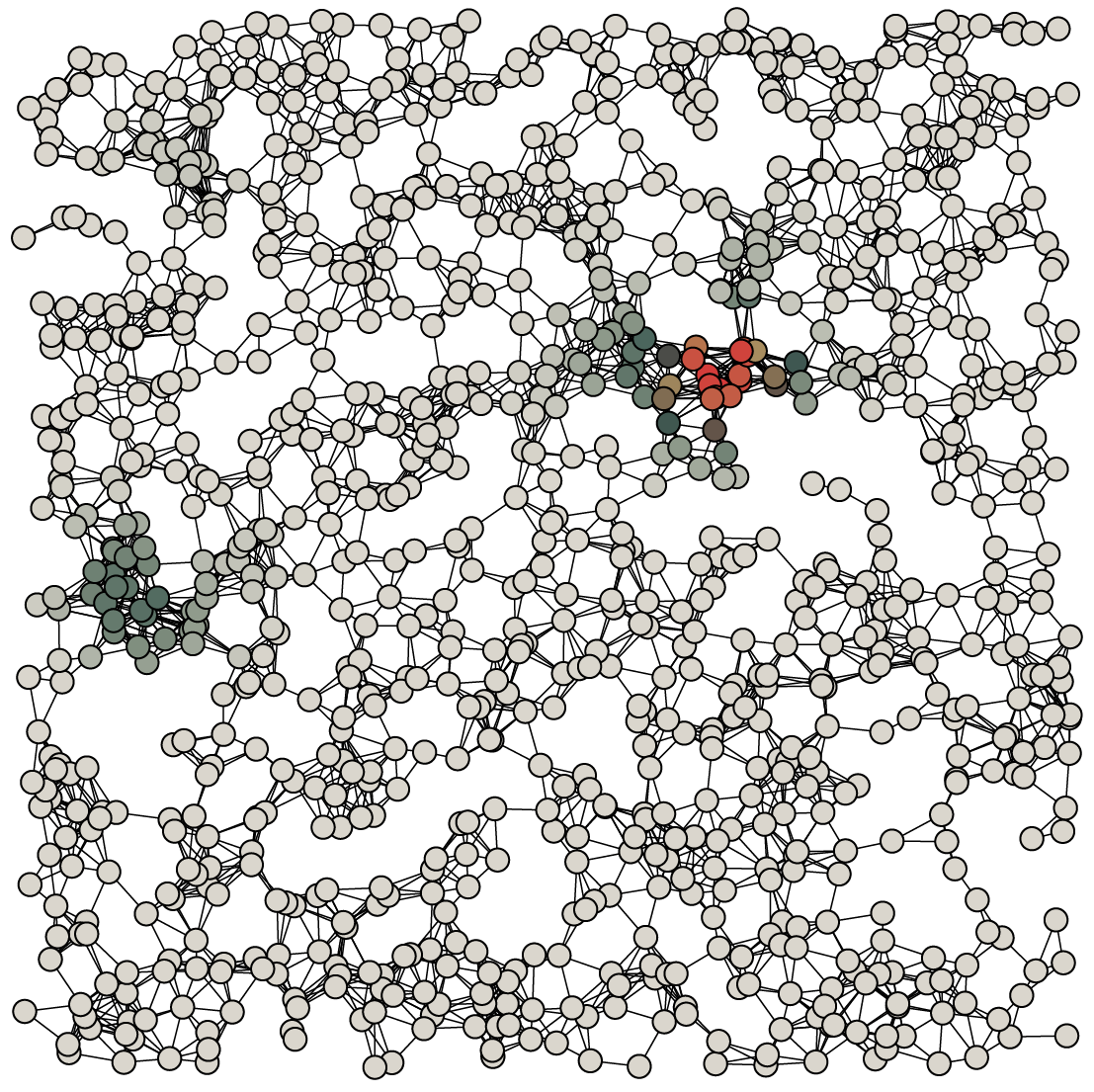}
	\caption{A random graph data structure with highlighted Katz centrality values.\protect\footnotemark~The more red a node is, the higher the centrality.}
	\label{fig:centrality}
\end{figure}\footnotetext{\url{https://upload.wikimedia.org/wikipedia/commons/9/9e/Wp-01.png}\label{footnote:wiki}}


\autoref{fig:centrality} shows an example of a random graph structure where Katz centrality was used to assign each node an appropriate score.
This method has some limitations; for instance, a Linux system likely does not have every available package installed.
Calculating the dependency graph and centrality scores for such a system might yield different values.
In this analysis, only the package repository with all nodes in their default configuration is considered.
No special settings that, for example, exclude particular libraries, were considered.


In order to create a dependency graph for an entire Linux distribution, the NixOS software repository, called nixpkgs, was chosen.
The advantage of this software repository compared to others like Debian or Fedora is its special design: each package is described as a function in the nix programming language.
The nix language is a functional language specifically designed for software packaging.
From a technical perspective, the entire nixpkgs repository is considered a large program.
Therefore, there is no need to crawl packages from the relevant repository or parse the embedded software package metadata.
Using the nix language, this information can be accessed directly with minimal sources of error.


To summarize the approach, a program in the nix language was written to traverse all software packages and their dependencies.
The traversed graph was then exported as a \gls{json} object and imported into a Python program for further analysis with the NetworkX library.
The NetworkX library was used to calculate the Katz centrality and sort the dependency graph.
The first 200 nodes of the sorted graph were considered for further analysis.

\begin{figure}[htb]
	\centering
	\includegraphics[width=1\linewidth]{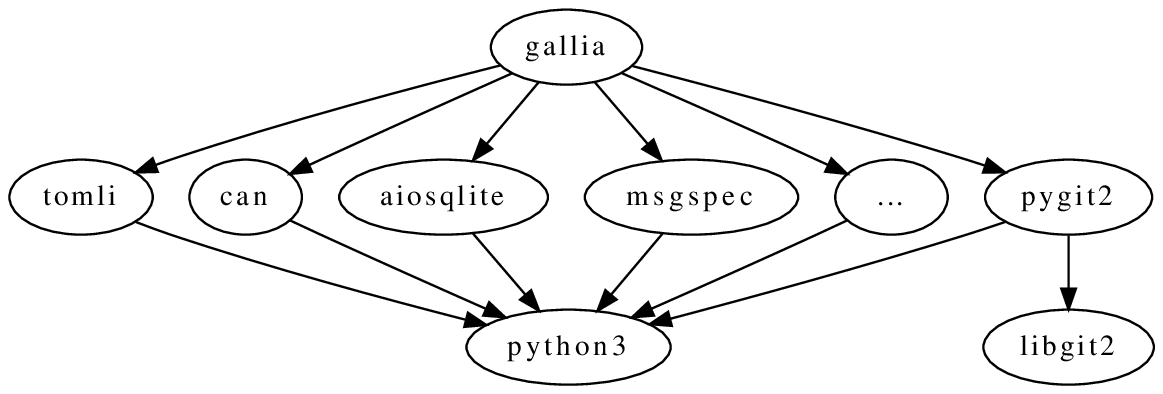}
	\caption{A subgraph of dependencies of the gallia \cite{Tatschner_gallia} package.}
	\label{fig:gallia-deps}
\end{figure}


A subgraph extracted from the created graph database is shown in \autoref{fig:gallia-deps}.
A software package is represented by a graph node, which can have edges labeled with \texttt{DEPENDS\_ON} to other nodes.
Not shown in the figure are properties that can be assigned to nodes, such as the package name, software version, or used licenses.


The canonical software repositories only include information for libraries in the language specific to the ecosystem.
For instance, the \gls{pypi} only provides information for Python dependencies.
System libraries like OpenSSL are not included in the dependency manifests provided by \gls{pypi}.
The created dependency graph in this study can even be queried for, e.g., system libraries that are required by Python libraries.

\subsection{Determination of Relevant Vulnerability Databases}
\label{sec:vulndbs}


A vulnerability database is a structured collection of information about security vulnerabilities in software and hardware systems, detailing aspects such as vulnerability ID, description, severity, and mitigation strategies. Examples include the \gls{nvd}\footnote{\url{https://nvd.nist.gov/}}, which is a U.S. government repository of standards-based vulnerability data, or the \gls{osv}\footnote{\url{https://osv.dev/}}, which focuses on vulnerabilities in open-source projects. These resources help organizations identify and address security weaknesses.

%

Various definitions of computer security vulnerabilities have been published, such as:

\begin{itemize}
	\item \textbf{RFC4949} \cite{rfc4949}: “A flaw or weakness in a system's design, implementation, or operation and management that could be exploited to violate the system's security policy.”
	\item \textbf{ISO 27005} \cite{iso27005}: “A weakness of an asset or group of assets that can be exploited by one or more threats, where an asset is anything that has value to the organization, its business operations, and their continuity, including information resources that support the organization's mission.”
	\item \textbf{Committee on National Security Systems (CNSS) Glossary}\footnote{{\url{https://www.niap-ccevs.org/Ref/CNSSI_4009.pdf}}}: “A known weakness in a system, system security procedures, internal controls, or implementation by which an actor or event may intentionally exploit or accidentally trigger the weakness to access, modify, or disrupt normal operations of a system-resulting in a security incident or a violation of the system's security policy.”
\end{itemize}


In cybersecurity, the primary vulnerability database is the \gls{nvd}\footnote{\url{https://nvd.nist.gov/}}, which is operated by the \gls{nist}.
Vulnerabilities listed in the \gls{nvd} are referred to as \gls{cve}.
Each \gls{cve} has a unique identifier, such as CVE-2014-0160, which is used to track the status (e.g., vulnerable or fixed) of this vulnerability in software distributions.


There are entities known as \glspl{cna} that are authorized with specific scopes and responsibilities to regularly assign \gls{cve} IDs and publish the corresponding \gls{cve} Records.
For example, the Linux Kernel organization\footnote{\url{https://kernel.org}}, the Python Software Foundation\footnote{\url{https://www.python.org}}, and the curl project\footnote{\url{https://curl.se}} were recently accepted as \glspl{cna} and are permitted to allocate \gls{cve} IDs for their managed projects.
The current Top-Level Root entities that can allocate \gls{cve} IDs are\footnote{\url{https://www.cve.org/PartnerInformation/ListofPartners}} the \gls{cisa} and the MITRE Corporation.


The \gls{nvd} has a software ecosystem-agnostic scope, meaning vulnerabilities identified in any software project can be submitted to the relevant \gls{cna}.
However, these vulnerabilities must be mapped to the relevant software packages in software distribution to be practically useful.
Various attempts are maintained by different organizations:


\begin{itemize}
	\item \textbf{Software Repositories}: Ecosystem-specific software repositories, such as \gls{pypi} (for Python) or pkg.go.dev (for Go), maintain their own databases that connect \gls{nvd} database entries with specific software packages. Additionally, the status of issues, such as vulnerable or patched, is tracked.
	\item \textbf{Linux Distributions}: Similar to ecosystem-specific repositories, several Linux distributions, such as Arch Linux, Debian, Fedora, or Gentoo, maintain separate security databases. These databases also link \gls{nvd} database entries to specific package names within the context of the Linux distribution.
	\item \textbf{Software Projects}: A few software projects maintain effective tracking of security issues themselves. Recently, there has been a move towards more \glspl{cna} where software projects implement security tracking and even assign their own \gls{cve} numbers.
	\item \textbf{Aggregated Databases}: Additionally, there are independent projects that collect data from all other security databases. Examples include the Github Advisory Database, osv.dev, or the vulnerability database provided by mend.io.
\end{itemize}


A problem that arises with the variety of different databases publishing different views of the same raw data is the use of a common data format.
There is an effort led by Google to unify the data structure of these databases\footnote{\url{https://ossf.github.io/osv-schema/}}.


Due to the absence of a security tracker in NixOS, the security database of the Debian project was considered for this analysis.
The Debian project provides a software repository of comparable size to NixOS.
Since Debian is a release-based Linux distribution, it was expected that the structure of the package repository would be in a state comparable to that of NixOS, which is also release-based.
The Debian security database offers an easy-to-use \gls{json}-based \gls{api}.

\subsection{Addition of Missing Data and Filtering}


The dependency graph contains a wealth of information, such as the package name, the software license, and the supported platforms. For this study, the \gls{url} of the Git repository, the programming language of the relevant software project, a category, and the backers were considered relevant.


Important information, such as the URL to the Git repository of the source code, is not present in the created graph data structure. The nixpkgs repository needs access to the relevant source code to build the software packages. For this purpose, the entire build step, which includes downloading the source code, is defined as a nix function that does not expose the \gls{url} of the Git repository. Additionally, signed tarballs or similar are often used instead of directly cloning the Git repository. Therefore, it is necessary to manually search for the source code repositories' \glspl{url} and add them to the dataset.


The same manual approach applies to identifying the programming language. A straightforward method for automating this process would be to identify the build system used. However, this approach is error-prone, as most build systems support multiple programming languages in various configurations. Another possibility would be to count the filename extensions of all source code files. This approach is also error-prone because projects can contain supplementary data (e.g., documentation), which would lead to inaccurate results. Since the categories were defined by the authors to gain an overview of the found software projects for informational purposes, these too had to be determined manually.


After adding missing data to the dataset manually, it became apparent that some entries could not be further evaluated.
For instance, there were duplicates due to different package versions (e.g., Python 3.10 and Python 3.11).
Surprisingly, a few projects still maintain their code in legacy \glspl{vcs}, such as the \gls{cvs}, with the most recent release from 2008. In total, 35 entries were filtered out from the 200 database entries.
The following analysis was conducted with 165 projects.

\subsection{Collection of Metrics}


The simplest approach to obtaining a meaningful number for a project's overall maintenance state would be to use the \gls{lma} value from \cite{COELHO2020106274}.
There are two limitations that prevent the use of the \gls{lma} in this study.
First, the authors did not publish their pre-trained model or code that could be used to create a setup of our own.
Second, the authors only considered repositories on Github.
Since many important software projects are hosted on different Git servers, such as those provided by the GNU project, the approach shown in \cite{COELHO2020106274} is not applicable to this study.


However, there are indications that some parameters are correlated with the \gls{lma} value, such as the number of core contributors (also known as the “bus factor”), lines of code, and commit activity.
Besides the \gls{lma}, there is currently no method available that can transform these parameters collectively into a number that can be used to compare the maintenance status of software projects.
Therefore, in this paper, only the collected data is presented and discussed.
All values are collected directly from the Git repositories using statistical methods and tools, such as pola.rs\footnote{\url{https://pola.rs/}}.

\subsection{Collection of CVEs}


Since the NixOS project does not offer a security tracker that maps actual \glspl{cve} to the NixOS definition of a package, a hybrid approach was decided upon.
The NixOS repository was used to create a complete dependency graph, and the Debian security database was used to obtain the actual state of security issues in those packages.
Obtaining this information was straightforward, as the Debian project provides a \gls{json}-based \gls{api} for the database.


The NixOS package names could be mapped to those used in Debian by manually searching the names of the source packages in Debian.
The search functionality was sufficient to find corresponding packages quickly.
However, this approach is prone to some errors:

\begin{itemize}
	\item \textbf{Human error}: Since the data is mapped by hand, some careless mistakes could occur. The authors double-checked the data to minimize these errors.
	\item \textbf{Missing packages}: Some packages available in NixOS were not available in Debian. Such packages are skipped in this analysis. Nine packages were not available in Debian at the time of writing.
	\item \textbf{Different splits}: Linux distributions tend to split packages to minimize the required disk space. For instance, a small program shipping a lot of documentation could be split into two packages: the program itself and the documentation. Splitting packages is a distribution-specific choice. By searching the Debian \emph{source} packages, most package splits are considered.
\end{itemize}

\section{Evaluation}
\subsection{Classification of Projects}

\begin{figure*}[htb]
	\begin{subfigure}[t]{0.45\textwidth}
		\includegraphics[width=\linewidth]{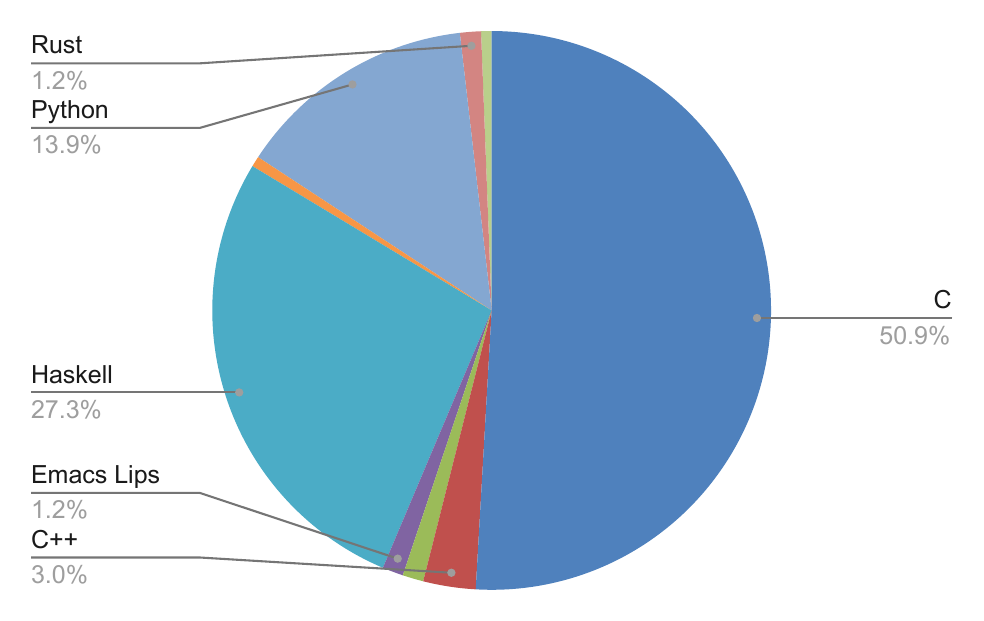}
		\caption{A pie chart showing the segmentation of used programming languages.}
		\label{fig:languages}
	\end{subfigure}\hfill
	\begin{subfigure}[t]{0.45\textwidth}
		\includegraphics[width=\linewidth]{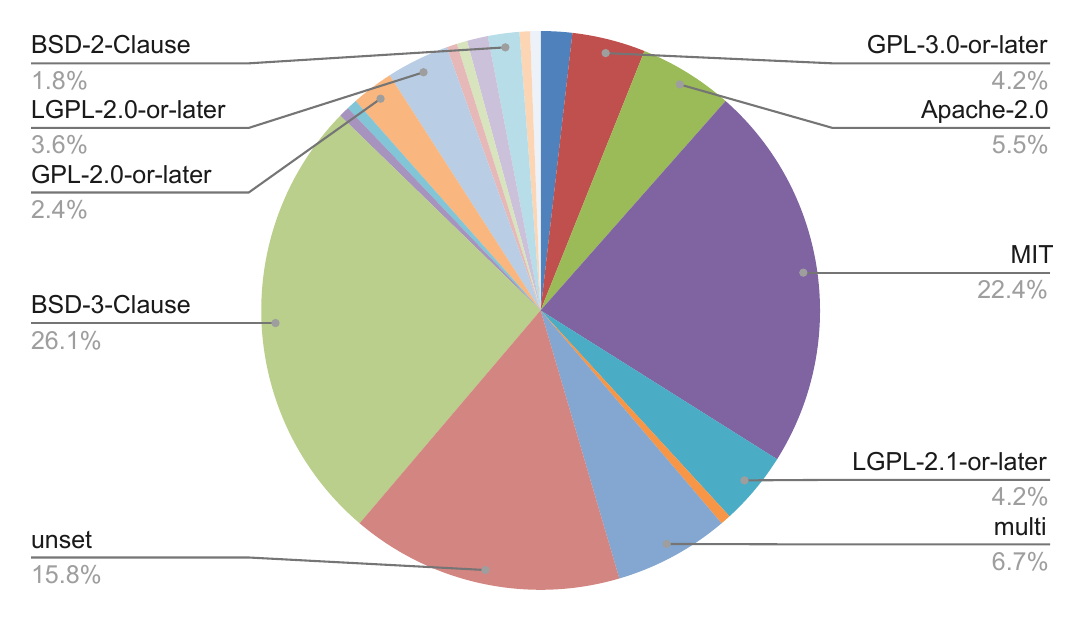}
		\caption{A pie chart showing the distribution of used licenses. The SPDX identifiers are used for license names. Multi licensed projects are noted with “multi”; projects with an empty license field are indicated with “unset”.}
		\label{fig:licences}
	\end{subfigure}
	\medskip
	\begin{subfigure}[t]{0.45\textwidth}
		\includegraphics[width=\linewidth]{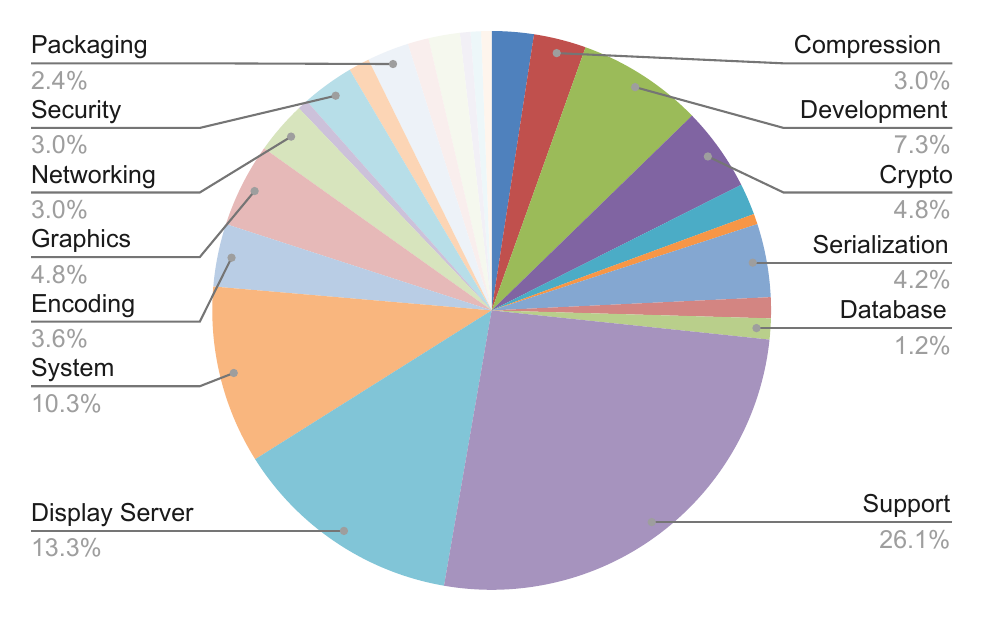}
		\caption{A pie chart showing the categories of the examined software projects. The categories have been defined and assigned in this work.}
		\label{fig:categories}
	\end{subfigure}\hfill
	\begin{subfigure}[t]{0.45\textwidth}
		\includegraphics[width=\linewidth]{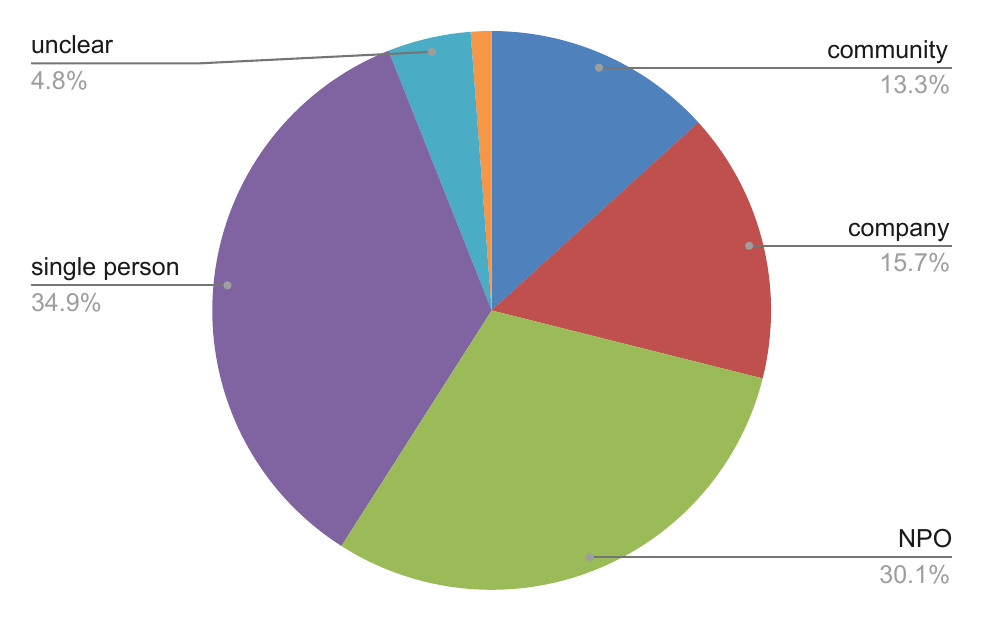}
		\caption{A pie chart showing the backers of the examined projects. NPO stands for Nonprofit organization.}
		\label{fig:backers}
	\end{subfigure}

	\caption{Four pie charts showing different analyses according to the dataset's classification.}
	\label{fig:three graphs}
\end{figure*}


A large dependency graph was created for all packages provided by the NixOS Linux distribution.
The created graph contains 82,011 nodes representing software packages and 273,681 edges representing dependency relations.


\autoref{fig:languages} shows the segmentation of used programming languages.
Most projects, more than half of all analyzed projects, are written in C.
The second most used language is Haskell, followed by Python, C++, and Rust.


\autoref{fig:licences} shows the used licenses in the dataset.
The most used licenses are BSD-3-Clause and MIT.
Some projects had an empty license field in the dataset, which is indicated with “unset”.
Some projects had multiple licenses.
Those cases were grouped together and indicated with “multi”.


\autoref{fig:categories} shows the categories defined and assigned by the authors.
These categories help to better understand the dataset and were used for sanity checking.
The most used category is “Support”, which is used to tag software libraries that only implement basic data structures, such as lists.
The second most used category is “Display Server”, which is used for all kinds of X-Server and Wayland-related libraries.


\autoref{fig:backers} shows the backers of the examined projects.
Most projects are maintained by a single person, with no affiliation to, for example, a company that could be determined.
The second most projects are backed by a \gls{npo}, such as the GNOME Foundation or GNU, followed by companies.

\subsection{Parameters of Projects}

\begin{figure}[htb]
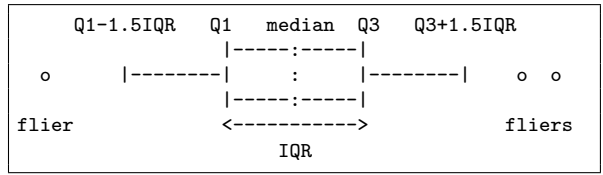

    \centering
\begin{Verbatim}[fontsize=\footnotesize,frame=single,samepage=true]
     Q1-1.5IQR   Q1   median  Q3   Q3+1.5IQR
                  |-----:-----|
  o      |--------|     :     |--------|    o  o
                  |-----:-----|
flier             <----------->            fliers
                       IQR
\end{Verbatim}
    \caption{Sketch showing the used configuration for box-and-whisker diagrams in this paper.}
    \label{fig:box-config}
\end{figure}

To better visualize statistical distributions, the following evaluations use box-and-whisker diagrams.
The configuration for all box-and-whisker diagrams, shown in \autoref{fig:box-config}, is as follows\footnote{\url{https://matplotlib.org/stable/api/_as_gen/matplotlib.pyplot.boxplot.html}}.

The box extends from the first quartile~(Q1) to the third quartile~(Q3) of the data, with a line at the median~$\tilde x$.
The whiskers extend from the box to the farthest data point lying within 1.5 times the \gls{iqr} from the box.
Flier points are those beyond the ends of the whiskers.
Fliers are indicated with a circle, \texttt{o}.
In terms of standard deviation~$\sigma$, the box extends to $\pm 0.6745 \sigma$.
Fliers are beyond $\pm 2.698 \sigma$.

\begin{figure}[htb]
	\includegraphics[width=\linewidth]{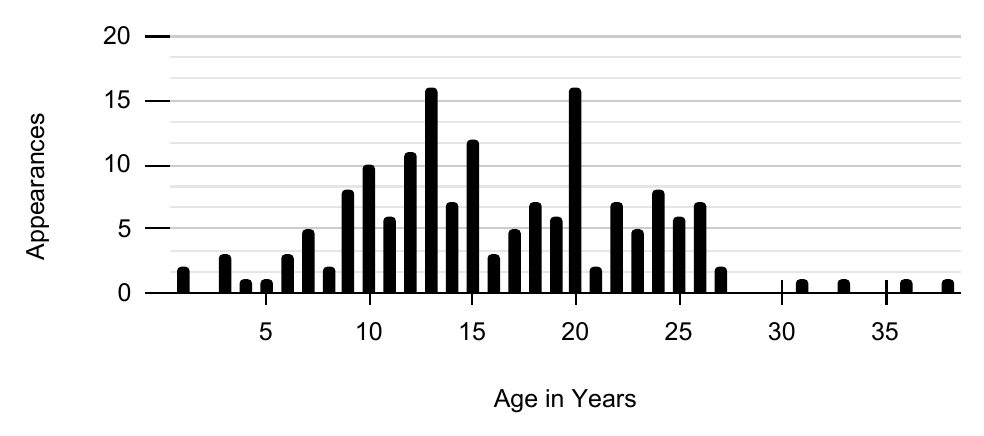}
	\caption{The age distribution of the chosen packages at the time of writing. The timestamp of the first commit in the Git repository is considered the start of the project.}
	\label{fig:age}
\end{figure}
\begin{figure}[htb]
	\includegraphics[width=\linewidth]{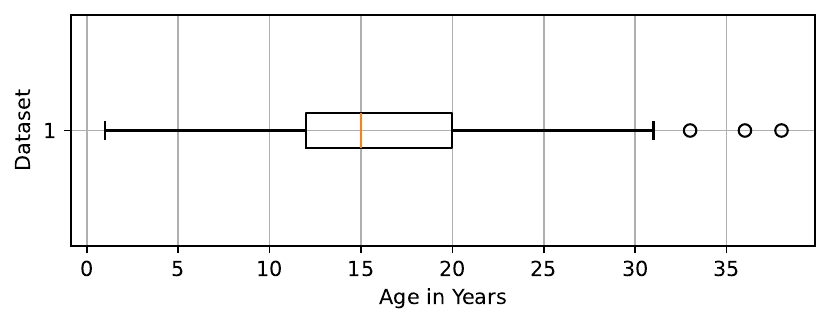}
	\caption{A box-and-whisker diagram of the project ages. The median is $\tilde x = 6$; projects older than 32 years are considered fliers.}
	\label{fig:ages-box}
\end{figure}


The authors defined the age of a project as the time difference between the first commit in the Git repository and the date of the evaluation (March 18, 2024).
\autoref{fig:age} shows the distribution of project ages across all analyzed projects.
Most projects considered important for the ecosystem are between 10 and 20 years old.
\autoref{fig:ages-box} shows a box-and-whisker diagram of the age distribution.
Projects with an age greater than 32 years are considered fliers in this evaluation.
These projects are: Python (33 years), Perl (36 years), and Emacs (38 years).

\begin{figure}[htb]
	\includegraphics[width=\linewidth]{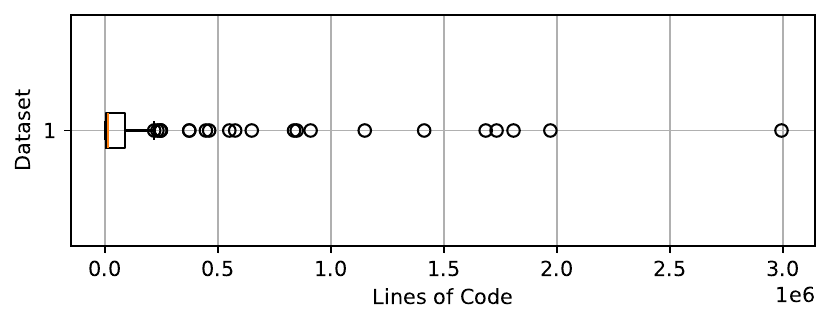}
	\caption{A box-and-whisker diagram of the number of lines of code. The median is $\tilde x = 12750$; projects with more than 2.2k lines of code are considered fliers. The fliers deviate by several orders of magnitude from the median.}
	\label{fig:locs-box}
\end{figure}

\autoref{fig:locs-box} shows the distribution of lines of code.
This value is widely spread.
The median is $\tilde x = 12750$ and the maximum observed value is 2,992,528.
The flier values deviate by several orders of magnitude from the median.
Projects which a high number of \gls{loc} are QT~(2,992,528~LoC), Emacs~(1,970,007~LoC), Python~(1,806,924~LoC) or Ruby~(1,731,991~LoC).

\begin{figure}[htb]
	\includegraphics[width=\linewidth]{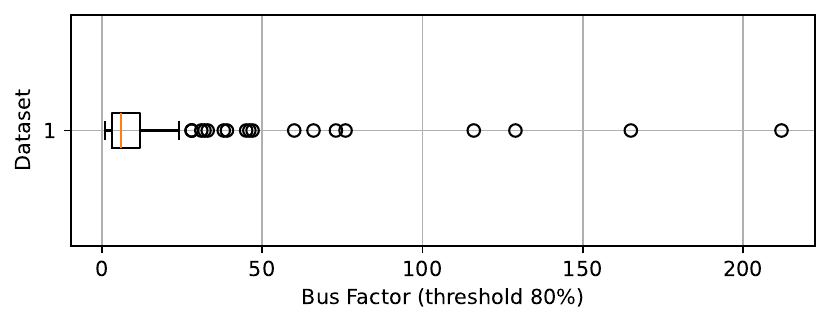}
	\caption{A box-and-whisker diagram of the bus factors. A bus factor of 1 in this analysis is defined as one author is responsible for 80\% of all commits. The median is $\tilde x = 6$. The largest observed value is 212.}
	\label{fig:bus-box}
\end{figure}

\autoref{fig:bus-box} shows the distribution of the bus factors.
In the literature, the “bus factor” is known as the minimum number of team members that have to suddenly disappear from a project before the project stalls due to lack of knowledgeable or competent personnel.
There are multiple different definitions how the bus factor can be calculated.
The authors decided to use the definition of number of core contributors from \cite{COELHO2020106274}, i.e., the number of authors who own more than 80\% of all commits.

The distribution looks similar as the distribution of lines of code in \autoref{fig:locs-box}.
However, the projects identified as fliers are different.
Projects with a remarkable high bus factor are:  gdk-pixbuf~(212), QT~(165), glib~(129), and at-spi2-core~(116).

\begin{figure}[htb]
	\includegraphics[width=\linewidth]{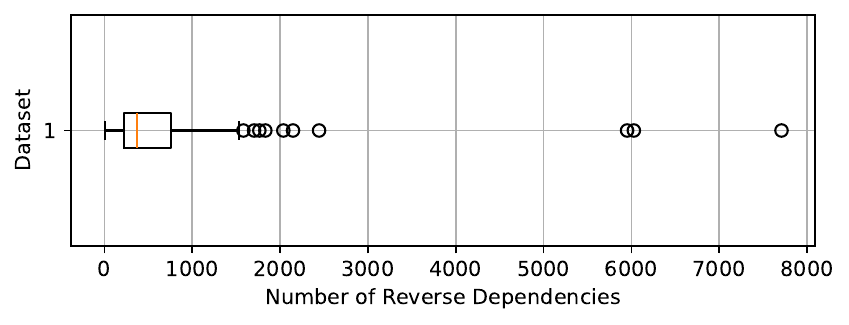}
	\caption{A box-and-whisker diagram of the reverse dependencies. The median is $\tilde x = 375$. The fliers deviate by one order of magnitude from the median. The maximum observed number is 7,709.}
	\label{fig:deps-box}
\end{figure}

\autoref{fig:deps-box} shows the distribution of the number of reverse dependencies.
The distribution looks familiar: a low median value with a small box and a few fliers that deviate by a magnitude from the median.
Projects with the highest number of reverse dependencies are Python~(7,709), Texinfo~(6,028), Emacs~(5,951), and Perl~(2,444).

\subsection{Open Issues}

The following charts combine the data from the dependency graph with the Debian security database.
As a first step, the Debian security database was examined.
In the database, there are 3,527 packages listed.
At the time of writing, there are 2,391 open issues where a \gls{cve} is considered not patched.

\begin{figure}[htb]
	\includegraphics[width=\linewidth]{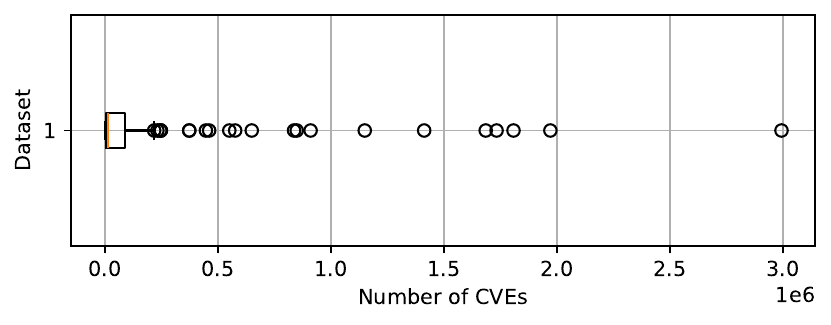}
	\caption{A box-and-whisker diagram of the overall number of \gls{cve} related entries in the Debian security database.}
	\label{fig:cve-box}
\end{figure}


\autoref{fig:cve-box} shows a box-and-whisker diagram of the number of overall security issues per package.
The package with the most entries in the database is the Linux kernel~(3,025 entries), followed by Chromium~(1,571 entries), Firefox~(1,347 entries), and Gitlab~(981 entries).
The median is $\tilde x = 2$ entries per package.

\begin{figure}[htb]
	\includegraphics[width=\linewidth]{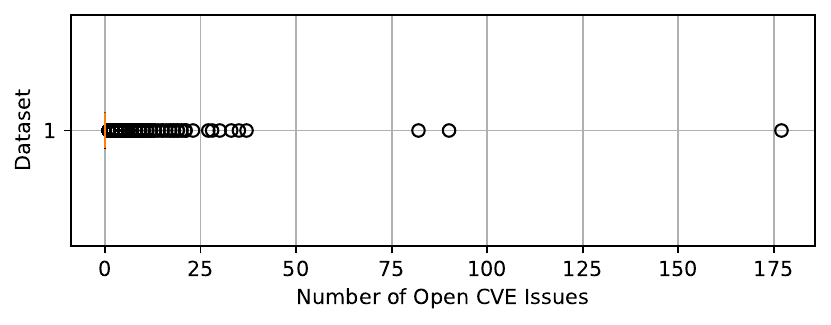}
	\caption{A box-and-whisker diagram of the unpatched number of \gls{cve} related entries in the Debian security database.}
	\label{fig:cve-box-unpatched}
\end{figure}

In addition, \autoref{fig:cve-box-unpatched} shows the number of unpatched \gls{cve} issues per package.
The values are widely spread.
The median is $\tilde x = 0$ and the box size is $\text{IQR} = 0$.
For the box-and-whisker diagram configuration used, all database entries with the number $n$ of open \gls{cve} issues where $n \neq \tilde x \neq 0$ are considered fliers.
The package with the most open issues is the Linux kernel~(177 entries), followed by TeX Live~(90 entries), gtkwave~(82 entries), and wpewebkit~(37 entries).

\begin{figure}[htb]
	\includegraphics[width=\linewidth]{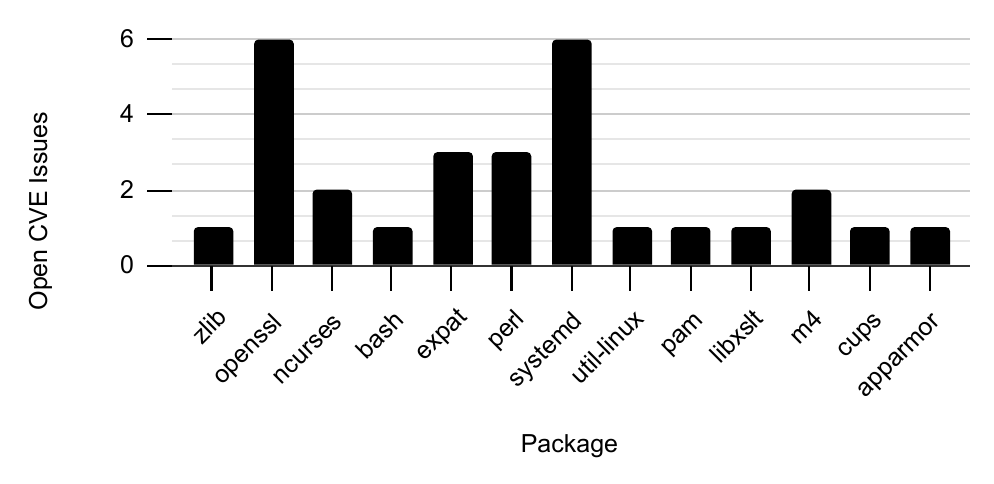}
	\caption{The number of open issues per package in Debian stable. The packages are ordered by the Katz centrality value from high values to lower values, where zlib has the highest value.}
	\label{fig:cves}
\end{figure}

In the chosen package list, there are 16 packages with open \gls{cve} issues, as shown by \autoref{fig:cves}.
The packages are ordered from the highest Katz centrality value to lower values, with zlib having the highest value.
Both systemd and OpenSSL have the highest number of unaddressed issues~(6 entries), followed by gdbm~(4 entries) and Perl/expat~(3 entries).

\begin{figure}[htb]
	\includegraphics[width=\linewidth]{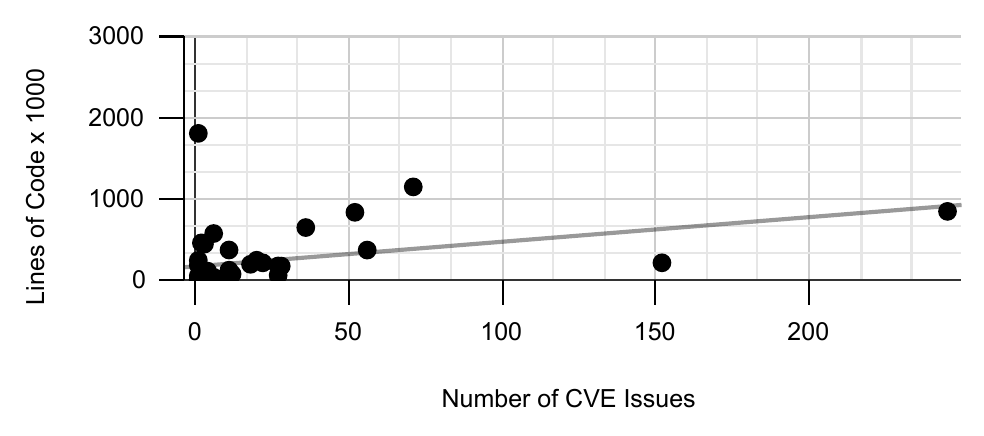}
	\caption{The number of \glspl{cve} in a particular package from the chosen dataset plotted against the number of lines of code. The linear regression line shows a possible correlation between the number of \glspl{cve} and \gls{loc}.}
	\label{fig:cve-loc}
\end{figure}

The scatter plot in \autoref{fig:cve-loc} shows the relationship between the overall number of \glspl{cve} in a particular package and the number of lines of code.
There seems to be a linear correlation between these.
Further analysis revealed that projects with a high number of \glspl{cve} are unmaintained projects where no developer is available to address the open issues.

\section{Discussion}

\subsection{Structure of the Ecosystem}


This study showed that the current \gls{foss} ecosystem suffers from multiple problems from an infrastructure point of view.
However, a software ecosystem consisting of 82,011 packages and 273,681 \texttt{DEPENDS\_ON} relations is already too large for a single human to grasp.
Since required metadata needed to be added manually (cf. \autoref{sec:methodology}), the presented approach is currently only feasible for a limited number of software packages.
For a large-scale analysis, the required interfaces, such as a standardized metadata format or a web-based \gls{api}, are not available in the ecosystem.
Furthermore, there is a lot of redundancy and uncertainty in the publicly available data hindering automated evaluation and monitoring.


For instance, the \gls{nvd} database serves as the de facto standard service that provides information regarding vulnerabilities.
Unfortunately, the \gls{nvd} does not provide standardized information about which particular software package in which particular software ecosystem is described.
To address this problem, there are multiple additional databases available (cf. \autoref{sec:vulndbs}) that utilize a standardized format developed by Google\footnote{\url{https://ossf.github.io/osv-schema/}}, which is not used by the \gls{nvd}, though.
The scope of these databases is always limited, e.g., to Github, a specific programming language ecosystem, or certain Linux distributions.
However, there is no vulnerability database available that is suitable for a large-scale analysis of the whole \gls{foss} ecosystem, including information about software dependencies.
In order to improve this situation, the development of an alternative software distribution architecture where the relationships between software modules are specified in a standardized cross-language and cross-ecosystem format is a topic for future research.


According to \autoref{fig:languages}, the Haskell programming language is the second most used language.
This was not expected, since Haskell is not commonly used for systems programming on Linux.
Further investigation using the proposed categories (cf. \autoref{fig:categories}) revealed that the Haskell software ecosystem is more separated than, e.g., the Python ecosystem.
In Haskell, there is no large “batteries included” standard library like the Python standard library\footnote{\url{https://docs.python.org/3/tutorial/stdlib.html\#batteries-included}}.
Instead, basic functionality, like data structure implementations, is maintained in separate software modules.
\autoref{fig:categories} shows a high occurrence of packages in the “Support” category.
Most of these packages are Haskell modules providing basic functionality, such as lists or dictionaries.
Consequently, Haskell modules appeared in the dependency graph with a high Katz centrality value.
Such separated ecosystems are very flexible, as new functionality can be added quickly.
However, it is not clear if the added complexity in terms of dependency tracking or developer resources is harmful to the Haskell ecosystem, especially compared to ecosystems like Python.


According to \autoref{fig:backers}, only 15.7\% of the examined projects' backers have an explicit company affiliation.
This is surprising considering the added value that the \gls{foss} ecosystems provide to a wide range of different companies.
Especially for critical projects, such as the ones identified by this paper, broader support from companies would be important to literally avoid single points of failure, e.g., maintenance by a single person as in the case of the xz Backdoor~(CVE-2024-3094).
Hence, both the general public and the companies themselves would benefit from a more resilient \gls{foss} ecosystem.


However, there might be the risk of companies actively pushing projects towards stricter or non-free licensing.
Recently, there was such a case in a popular open-source key-value database, Redis\footnote{\url{https://redis.com/blog/redis-adopts-dual-source-available-licensing/}}.
Due to an active \gls{cla}, the backing company changed the license of the project to a non-free alternative.
Consequently, developers left the project and created a fork\footnote{\url{https://github.com/valkey-io/valkey}}.
A further example is the Intel-backed Hyperscan library, which was recently converted into closed-source software\footnote{\url{https://www.phoronix.com/news/Intel-Hyperscan-Now-Proprietary}}.
The authors assume that it is only a matter of time until forks will start to appear.
Unfortunately, such situations increase the complexity of the open-source ecosystem rather than decreasing it.
This is especially worth mentioning since a majority of the analyzed projects use permissive license models such as BSD and MIT, as indicated by \autoref{fig:licences}.


The box-and-whisker diagrams in \autoref{fig:locs-box}, \autoref{fig:bus-box}, and \autoref{fig:deps-box} altogether show a common property of the examined software packages.
On the one hand, there are large and actively maintained projects by a community.
On the other hand, there are projects that are maintained by a small number of developers (i.e., a low bus factor).


At the time of writing, there were 2,391 open \gls{cve} entries in the Debian security database.
Using the presented Katz centrality-based method, it is possible to prioritize and sort those entries (cf. \autoref{fig:cve-box-unpatched}).

\subsection{CVE Handling}


From the analyzed projects, the curl project stands out as a good example of well-maintained software.
The curl project provides an implementation of a complete \gls{http} stack and has 663 reverse dependencies in NixOS.
Daniel Stenberg, the main maintainer, makes every effort to improve the project.
For instance, the project has recently been accepted\footnote{\url{https://daniel.haxx.se/blog/2024/01/16/curl-is-a-cna/}} as a \gls{cna} in order to issue its own \glspl{cve} entries.
Further, the project provides a dashboard\footnote{\url{https://curl.se/dashboard.html}} where multiple development metrics are tracked over time.
Interesting metrics include the “\gls{cve} age in code until fixed” and “curl vulnerabilities: C vs non-C mistakes.”
Such metrics are helpful to track the project state over time and identify emerging issues early.


The \gls{cve} infrastructure with the \gls{nvd} has received negative feedback in the past.
For instance, for curl, CVE-2020-19909 was filed and graded as a 9.8 CRITICAL issue\footnote{\url{https://daniel.haxx.se/blog/2023/08/26/cve-2020-19909-is-everything-that-is-wrong-with-cves/}}.
It turned out that it was indeed a bug but with no security implications.
However, the alarmism spread, and curl was tagged as insecure by Linux distributions.
MITRE rejected Daniel Stenberg's requests to withdraw the faulty \gls{cve} due to the existence of a valid weakness (integer overflow), which could result in a valid security impact\footnote{\url{https://curl.se/docs/CVE-2020-19909.html}}.


After further discussion, the \gls{cve} was eventually re-scored as 3.3, and curl was accepted as a \gls{cna}.
The main point of criticism in this situation was that any person can file \glspl{cve}.
There is no verification process in place where, for example, a maintainer must approve that it is indeed security-relevant.


Greg Kroah-Hartman, a kernel maintainer who is sponsored by the Linux Foundation, also talked about the \gls{nvd} situation in the past\footnote{\url{https://kernel-recipes.org/en/2019/talks/cves-are-dead-long-live-the-cve/}}.
He describes a similar situation with CVE-2019-12357, where a faulty \gls{cve} was assigned and caused repercussions.
It turned out that patching the said \gls{cve} was not necessary, and the change was eventually reverted.
In software companies, addressing security-related issues usually has a higher priority than minor bug fixes; hence, it is easier to justify developer resources when an appropriate \gls{cve} is available.
As a consequence, the Linux kernel community became a \gls{cna} in order to avoid faulty \glspl{cve} in the future.

\subsection{Limitations}

The presented approach has several limitations that must be acknowledged to provide context for the study's findings and to guide future research efforts.

\begin{enumerate}
	\item \textbf{Number of analyzed packages}:
	      In the dataset of approximately 80,000 records, an upper limit of 200 records has been set for manual analysis.
	      This decision is based on balancing the need for detailed insights with the practical constraints of manual review, which are resource-intensive and time-consuming.
	      This manageable number allows deeper analysis of each record, providing richer and more nuanced insights.
	      Although 200 records do not suffice for comprehensive analysis, valuable insights can still be yielded if they are chosen to reflect the dataset's diversity and key characteristics.
	      This approach ensures a balance is struck between detail and practicality within the constraints of manual analysis.
	\item \textbf{Unidimensional Analysis}:
	      This paper examines projects not limited to those hosted on GitHub, which presents certain challenges for conducting a multi-dimensional analysis, including aspects like community activity.
	      Many projects, such as the Linux kernel, Git, or GNU projects, do not utilize GitHub for their community interactions.
	      Consequently, the tools and techniques from the \gls{ossf}, which are tailored specifically for GitHub-hosted projects, are not applicable to this study.
	      This limitation poses a problem for academic research, restricting the scope of available data and thus the comprehensiveness of the analysis in this paper.
	\item \textbf{Human Error in Manual Analysis}:
	      The lack of programming interfaces required a manual analysis approach, which presents a notable limitation.
	      Manual analysis is prone to human error, as it relies on individual judgment and execution, potentially leading to inconsistencies and inaccuracies.
	      While this approach allows for nuanced examination, the potential for human error remains significant, underscoring the need for automated solutions in future research to enhance result reliability and validity.
	\item \textbf{Ambiguous Mapping of Packages}:
	      A notable limitation of this study is the challenge in mapping package names from a specific software repository to entries in the corresponding vulnerability database.
	      This difficulty arises due to the lack of normalization in package names, leading to ambiguities and inconsistencies.
	      As a result, the process of accurately linking packages to their associated vulnerabilities is hindered, potentially affecting the reliability of our analysis.
	      Future work could focus on developing a standardized naming convention or implementing advanced algorithms to improve the accuracy of these mappings.
\end{enumerate}

The limitations identified in this study present opportunities for further research and future work to address and explore these challenges in greater depth.

\section{Conclusion}


This paper addressed the current challenges of managing large software repositories from a security point of view.
The \gls{foss} ecosystem nowadays consists of multiple different and interconnected sub-ecosystems.
Due to upcoming legal regulations, such as the European \gls{cra}, techniques to analyze and track the state of a software project and software ecosystems in general, with \glspl{sbom}, become more and more relevant.


In this study, it became apparent that the \gls{foss} ecosystem currently suffers from different problems, such as insufficient funding, companies that monetize projects by changing licenses to non-free alternatives, or faulty \gls{cve} reports for justifying developer resources.
From a technical point of view, there are also multiple problems.
Especially the lack of standardized interfaces will become a burden when implementing the new legal requirements at a large scale.


Considering the formulated research question \textbf{RQ1} \RQ, the authors concluded that sorting software modules by their impact is possible with a centrality-based approach.
Consequently, for this purpose the software modules have to be available in a dependency graph data structure.
The standardization of common interfaces for creating \gls{foss} ecosystem-wide examinations will be in the focus of future research.
The author postulates that the existence of such interfaces and dedicated automated evaluations helps to better detect supply chain attacks, such as the xz backdoor, early.


Considering the more general research question \textbf{RQ2} \RQQ, the authors tend to a two-minded answer.
On the one hand, there are well-maintained and well-funded projects such as Python, which have a high impact.
On the other hand, there are projects with a high impact but with a low bus factor that are vulnerable to supply chain attacks, as shown by the xz Backdoor.
This paper has contributed insights and methods to identify such critical projects.
This allows the community, and especially companies, to benefit from the \gls{foss} ecosystem by supporting these projects and hopefully preventing incidents such as the xz Backdoor in the future.

\section*{Data Availability}

The raw data of this study is available under the CC BY 4.0 Legal Code license at Zenodo \cite{tatschner_2024_11276931}.

\section*{Declaration of generative AI and AI-assisted technologies in the writing process}

During the preparation of this work the authors used ChatGPT in order to improve the language of the paper.
After using this tool, the authors reviewed and edited the content as needed and take full responsibility for the content of the published article.

\section*{Acknowledgments}


Many thanks to David Emeis and Veronique Ehmes for the productive technical discussions regarding vulnerability databases.
The authors would like to thank the Wikipedia user Pholme for publishing the image\textsuperscript{\ref{footnote:wiki}} used as \autoref{fig:centrality} under the CC BY-SA 4.0 Deed license.

\section*{Funding}

This work was supported by the German Federal Ministry of Education and Research (BMBF) under Grant No. 16KIS1847, ALPAKA.

\bibliographystyle{elsarticle-num}
\bibliography{bibliography}

\end{document}